\documentclass[sn-mathphys-num,iicol]{sn-jnl}

\usepackage{graphicx}%
\usepackage{multirow}%
\usepackage{amsmath,amssymb,amsfonts}%
\usepackage{amsthm}%
\usepackage{mathrsfs}%
\usepackage[title]{appendix}%
\usepackage{xcolor}%
\usepackage{textcomp}%
\usepackage{manyfoot}%
\usepackage{booktabs}%
\usepackage{algorithm}%
\usepackage{algorithmicx}%
\usepackage{algpseudocode}%
\usepackage{listings}%

\usepackage{bm}
\usepackage{CJK}
\usepackage{makecell}
\usepackage{lineno}

\theoremstyle{thmstyleone}%
\theoremstyle{thmstyletwo}%

\theoremstyle{thmstylethree}%

\begin{document}

\title[Article Title]{MolMetaLM: a Physicochemical Knowledge-Guided Molecular Meta Language Model}


\author[1,2]{\fnm{Yifan} \sur{Wu}}\email{wuyifan@csu.edu.cn}

\author[1]{\fnm{Min} \sur{Zeng}}\email{zengmin@csu.edu.cn}

\author[2]{\fnm{Yang} \sur{Li}}\email{liyangum@nus.edu.sg}

\author*[2]{\fnm{Yang} \sur{Zhang}}\email{zhang@nus.edu.sg}

\author*[1]{\fnm{Min} \sur{Li}}\email{limin@csu.edu.cn}


\affil[1]{\orgdiv{School of Computer Science and Engineering}, \orgname{Central South University}, \orgaddress{\street{No.932 South Lushan Road}, \city{Changsha}, \postcode{410083}, \state{Hunan}, \country{China}}}

\affil[2]{\orgdiv{Cancer Science Institute}, \orgname{National University of Singapore}, \orgaddress{\street{14 Medical Drive}, \postcode{117599}, \country{Singapore}}}



\abstract{Most current molecular language models transfer the masked language model or image-text generation model from natural language processing to molecular field. However, molecules are not solely characterized by atom/bond symbols; they encapsulate important physical/chemical properties. Moreover, normal language models bring grammar rules that are irrelevant for understanding molecules. In this study, we propose a novel physicochemical knowledge-guided molecular meta language framework——MolMetaLM. We design a molecule-specialized meta language paradigm, formatted as multiple $<\mathrm{S},\mathrm{P},\mathrm{O}>$ (subject, predicate, object) knowledge triples sharing the same $\mathrm{S}$ (i.e., molecule) to enhance learning the semantic relationships between physicochemical knowledge and molecules. By introducing different molecular knowledge and noises, the meta language paradigm generates tens of thousands of pretraining tasks. By recovering the token/sequence/order-level noises, MolMetaLM exhibits proficiency in large-scale benchmark evaluations involving property prediction, molecule generation, conformation inference, and molecular optimization. Through MolMetaLM, we offer a new insight for designing language models. }




\maketitle

\section{Introduction}

The development of deep learning, especially Language Models (LMs), has significantly boosted the fields of drug discovery, improving the accuracy and efficiency of downstream tasks such as molecule generation, optimization and property prediction. Currently, molecular LMs can be roughly divided into two categories. The first category is Masked Language Model (MLM) \cite{kenton2019bert}-based LMs, which leverage masked molecular information as input and yield reconstructed molecules as output \cite{mendez2021geometric,chithrananda2020chemberta,zhou2022uni}. However, MLM-based molecular LMs have several limitations: 1) they only focus on discriminative abilities, ignoring generative abilities; 2) different molecular information have different input forms, making it difficult to unify different pre-training tasks into one model; 3) they only consider the topological patterns formed by the atoms or bonds, disregarding the intrinsic physical or chemical properties associated with these patterns. The second category is Generative Language Model (GLM) \cite{radford2019language}-based LMs that align molecular linguistic descriptions to molecules and generate target sequences based on source sequences \cite{edwards2022translation,zhang2024chemllm}. Although GLM-based molecular LMs provide unstructured linguistics inputs and outputs that are more human-friendly and facilitate the integration of different tasks, these inputs or outputs are not necessary to solve chemical molecular tasks. Furthermore, they may interfere with the model's understanding of chemical and molecular knowledge. Additionally, general Large Language Models (LLMs) \cite{du2022glm,touvron2023llama,ouyang2022training} have demonstrated remarkable potential in the molecular field and can even outperform specialized MLM/GLM-based molecular LMs when equipped with suitably designed prompts \cite{molregpt}. However, training such LLMs is extremely time-consuming, labor-intensive, and costly. 

In order to establish a universal molecular LM framework that can address the above limitations, we propose the concept of molecular meta language. Meta language is defined as a language used to describe or represent the language itself \cite{jakobson1976metalanguage}. One example is the $<\mathrm{S},\mathrm{P},\mathrm{O}>$ triple in natural language, where $S,O$ represent subject and object entities, $P$ denotes the predicate relation from the subject to the object. Almost all natural languages can be represented as the triple and the well-known knowledge graph \cite{ehrlinger2016towards} also applied it for knowledge representation. Compared to natural language, meta language can express complex logical relationships between concepts and entities more precisely due to its fixed linguistic and semantic rules. It eliminates the ambiguity and fuzziness caused by individual biases present in natural language. This inspired us to design a molecular meta language framework for molecular LMs, enabling the model to focus more on learning complex relationships within molecular knowledge itself. We summary two key challenges in current molecular LMs: 1) how to unify different molecular tasks and represent them into a general language paradigm; 2) how to design a universal pre-training paradigm for multiple downstream tasks. From the perspective of the molecular meta language, we address the first challenge by designing the molecular meta language as a physicochemical knowledge-guided $<\mathrm{S},\mathrm{P},\mathrm{O}>$ triple. Here, molecules are assigned to $S$, physicochemical properties or labels from different downstream tasks are assigned to $O$, and property or task identities are assigned to $P$, thereby unifying different molecule-related tasks. For the second challenge, we extend the denoising pre-training paradigm \cite{lewis2020bart,tay2022ul2} at the molecular level, which unifies MLM/GLM-based molecular LMs. The denoising pre-training paradigm views MLM, GLM as a denoising process from a high-level perspective. In this process, noises are introduced to the input sequence by masking tokens \cite{kenton2019bert,cui2021pre}, replacing them with synonyms \cite{cui2020revisiting}, removing key sentences \cite{luo2022gap}, shuffling the order of tokens \cite{yang2019xlnet}. The model is then trained to recover the noised sequences. By integrating the two ideas in molecular LMs, we propose a powerful universal molecular meta language framework called MolMetaLM. 

\begin{figure*}
    \centering
    \includegraphics[scale=0.1, trim=0 0 0 0]{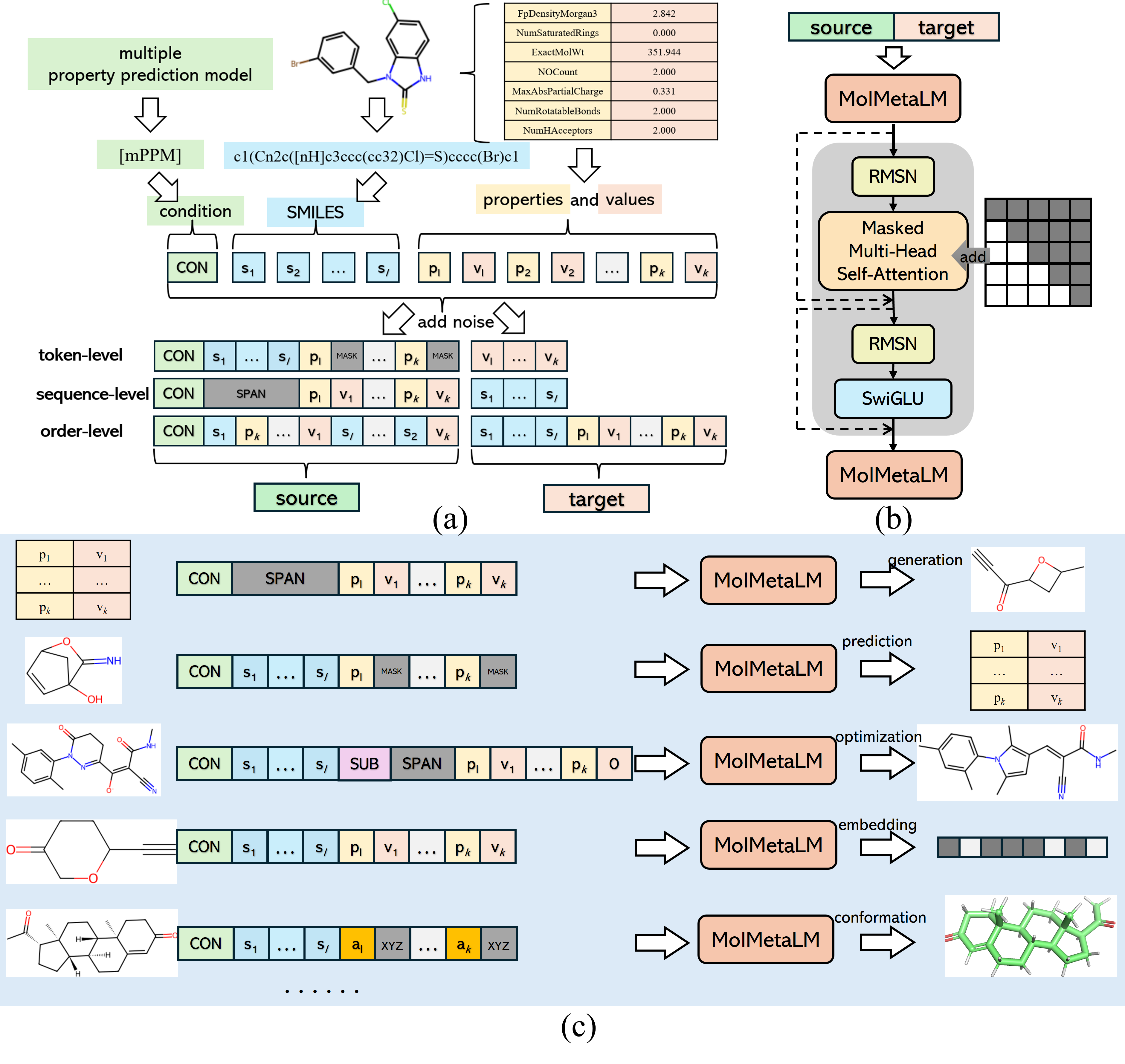}
    \caption{General framework of MolMetaLM. (a) Construction of the input meta language. The input to MolMetaLM is a mixture of k $<\mathrm{S},\mathrm{P},\mathrm{O}>$ triples that share the same $\mathrm{S}$. Then the token-level, sequence-level, and order-level noises are added to the input to construct the source and target sequences. The noise design at different level drives the model to learn to handle different denoising scenarios, so as to achieve the generation goal in different tasks. (b) The backbone of MolMetaLM. It is a stacked transformer decoder variant that applies the RMSNorm, SwiGLU, and rotary position embedding. (c) The application of MolMetaLM. With different design of the input meta language, MolMetaLM demonstrates proficiency in molecular generation, molecular optimization, property prediction, structure prediction, and other tasks.}
    \label{general_framework}
\end{figure*}

In MolMetaLM, we consider physicochemical property prediction tasks, fingerprint prediction tasks and conformation prediction tasks as the construction of $P$ and $O$. As shown in Figure \ref{general_framework}, the input meta language sequence is designed as: 
\begin{equation}
s_1,s_2,...,s_l, p_1,v_1,p_2,v_2,...,p_k,v_k
\end{equation}
\noindent where $s_1,...,s_l$ represent the tokens in the molecular SMILES \cite{weininger1988smiles}, $p_1,...,p_k$ denote property names, $v_1,...,v_k$ represent property values. It is a mixture of $k$ $<\mathrm{S},\mathrm{P},\mathrm{O}>$ triples that share the same $\mathrm{S}$, i.e., $<\underbrace{s_1,...,s_l}_{\mathrm{S}},\underbrace{p_1}_{\mathrm{P}_1},\underbrace{v_1}_{\mathrm{O}_1}>$, ..., $<\underbrace{s_1,...,s_l}_{\mathrm{S}},\underbrace{p_1}_{\mathrm{P}_k},\underbrace{v_1}_{\mathrm{O}_k}>$. The noises are defined as three types: 1) token-level noise; 2) sequence-level noise; 3) order-level noise. By adding the token-level noise, the patterns formed by atoms or bonds are captured to enhance the molecular representation abilities. The sequence-level noise mainly corresponds to the generation ability that enables the model to generate molecular SMILES based on given property conditions. The order-level noise is implemented by shuffling the input sequence, driving the model to reorder tokens into a valid SMILES representation and establish accurate correspondences between property names and property values. To establish a comprehensive and universal molecular LM framework, we design 18 basic denoising pre-training tasks within the molecular meta language, covering molecular physicochemical properties, molecular fingerprints, and conformations. These tasks are incorporated into the training of MolMetaLM. In large-scale benchmark evaluations, MolMetaLM demonstrates remarkable performance, underscoring its exceptional capabilities and versatility in the field of molecular generation and property prediction. 

\section{Results}

\subsection{Assessment of generation ability}

\begin{figure*}
    \centering
    \includegraphics[scale=0.43, trim=0 0 0 0]{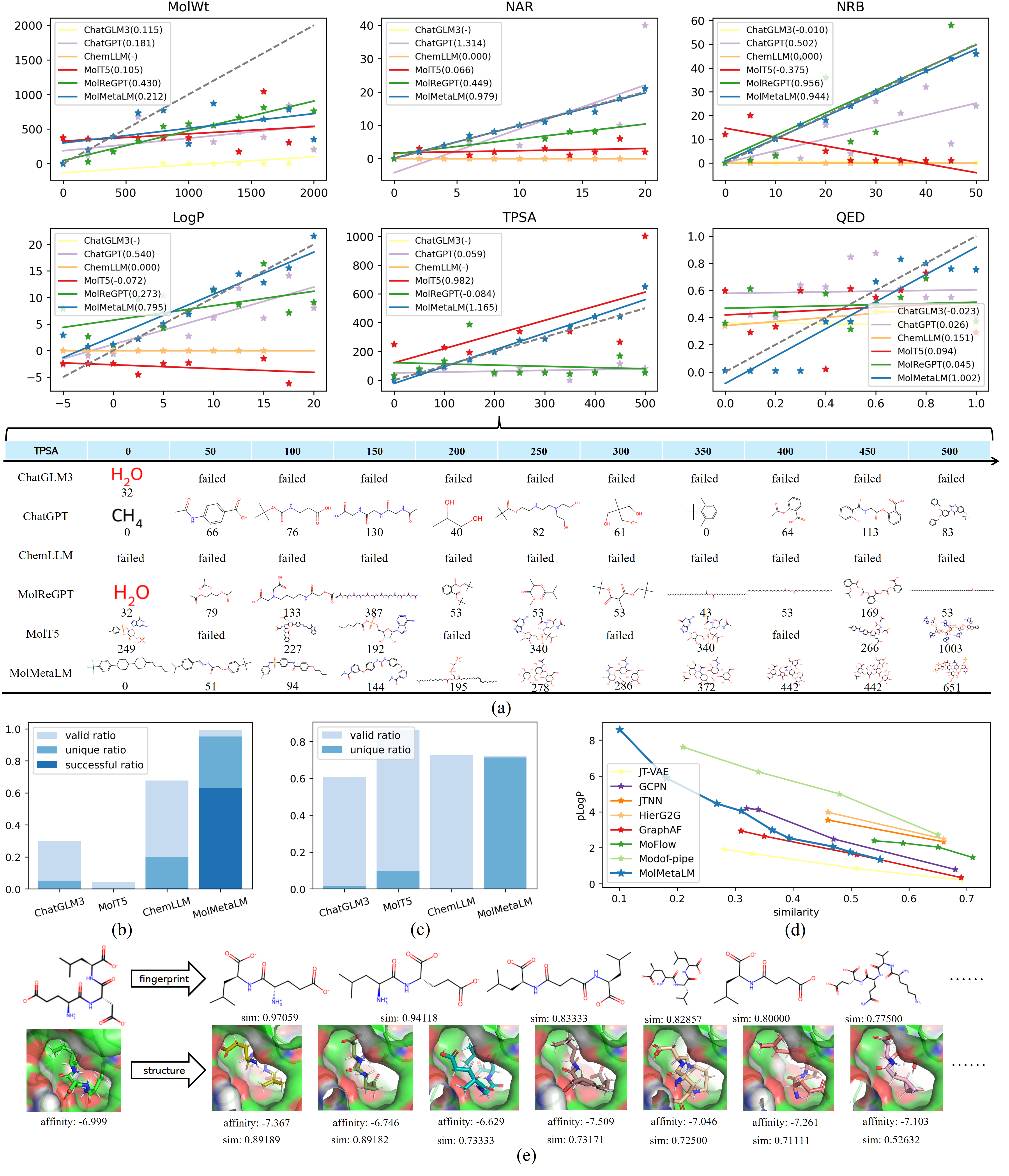}
    \caption{(a) The relation between the conditional property values and the values of the generated molecules from different methods. For each sub-figure of the upper part, the x-axis indicates the given property value constraints, y-axis indicates the property values of the generated molecules. For each method, if the generated molecules are valid, the corresponding points will be marked, and all marked points are fitted as a curve of degree one. The slopes of the curves are included in parentheses in the legends. It is worth noting that a slope closer to 1.0 does not always indicate better performance. For instance, in the generation conditioned by TPSA (bottom part), despite MolT5 exhibiting a slope of 0.982, it struggles to generate molecules with the given TPSA value and instead generates molecules with a value of 1003 in the generation task of TPSA 500, resulting in a slope close to 1.0. (b) Results of multiple-condition molecule generation. (c) Results of unconditional molecule generation. (d) Curves of similarity and pLogP improvement on Jin's test set. (e) Fingerprint/structure-based molecule generation cases using MolMetaLM. For fingerprint-based molecule generation, it generates molecules with high Tanimoto similarity based on MACCS; for structure-based molecule generation, it not only generates similar molecules but also generates ones with similar high docking affinities. The docking affinities are obtained from AutoDock Vina. }
    \label{all_res_generation}
\end{figure*}

\noindent\textbf{Single-condition molecule generation.} We mainly focus on generating molecules conditioned by QED (Quantitative Estimate of Drug-likeness), LogP (Partition coefficient), MolWt (Molecular Weight), TPSA (Topological Polar Surface Area), NRB (Number of Rotatable Bonds) and NAR (Number of Aromatic Rings). Figure. \ref{all_res_generation} (a) presents scatter plots of property values generated by different methods under different property conditions. ChatGPT-3.5 \cite{ouyang2022training}, as a general language model, even performs better than other chemistry/molecule-specific language model (ChemLLM \cite{zhang2024chemllm}, MolT5 \cite{edwards2022translation}) in certain conditions. Both ChatGLM3 \cite{du2022glm} and ChemLLM generate many invalid SMILES. We speculate this issue is caused by their training corpuses. ChatGLM3 focuses more on daily chat, while ChemLLM focuses more on molecular synthesis or description generation. The most difficult thing for language models is how to understand the differences between different numerical values at the text level. Benefiting form the designed molecular meta language, MolMetaLM can better focus on learning the relationships between molecules and property entities and understand the given property value constraints from the text level, leading to the generation of more appropriate molecules.

\noindent\textbf{Multiple-condition molecule generation.} In order to simulate practical application scenario in molecule design, we define the conditional constraints based on Lipinski's five rules \cite{lipinski2012experimental}. Specifically, we constrain MolWt to random numbers within the range of 0 to 500, NHD (Number of Hydrogen bond Donors) within 0 to 5, NHA (Number of Hydrogen bond Acceptors) within 0 to 10, LogP within -2 to 5, and NRB within 0 to 10. The results in terms of valid ratio, unique ratio and successful ratio are reported in Figure. \ref{all_res_generation} (b). MolMetaLM shows superior performance compared to other methods, with over 90\% for both valid ratio and unique ratio, as well as over 60\% for the successful ratio (defined as the ratio of unique generated molecules satisfying the Lipinski's five rules). The results align with our expectations, because ChatGLM3, MolT5, and ChemLLM are all biased by non-binding natural language. In addition, conditioned by fingerprint bits or reference molecular backbone coordinates, MolMetaLM can also generate molecules that are similar to the input molecules (Figure. \ref{all_res_generation} (e)), enabling MolMetaLM has great potential in practical drug design.

\noindent\textbf{Unconditional molecule generation.} Unconditional generation can reflect the diversity of model generation space. MolMetaLM is compared with the existing pure language models (Figure. \ref{all_res_generation} (c)). It can be observed that although MolT5 exhibits the highest valid ration, it generates duplicated molecules which results in a poor unique ratio. Similar issues are observed in ChatGLM3 and ChemLLM. In contrast, MolMetaLM achieves almost 100\% unique ratio, demonstrating its powerful ability to generate diverse molecules.  

\noindent\textbf{Molecular property optimization.} Following Jin et al. \cite{jin2018junction}, we focus on the optimization of pLogP (penalized LogP) which measures the partition coefficients penalized by synthetic accessibility and ring size. Without any further fine-tuning, the performance of MolMetaLM on Jin's test set \cite{jin2018learning} is shown in Figure. \ref{all_res_generation} (d). MolMetaLM achieves acceptable results compared to the previous methods, which are specifically well-trained or designed models for molecular optimization. This achievement underscores the ability of MolMetaLM in the domain of molecular property optimization. 

\subsection{Assessment of prediction ability}

\begin{figure*}
    \centering
    \includegraphics[scale=0.1, trim=0 0 0 0]{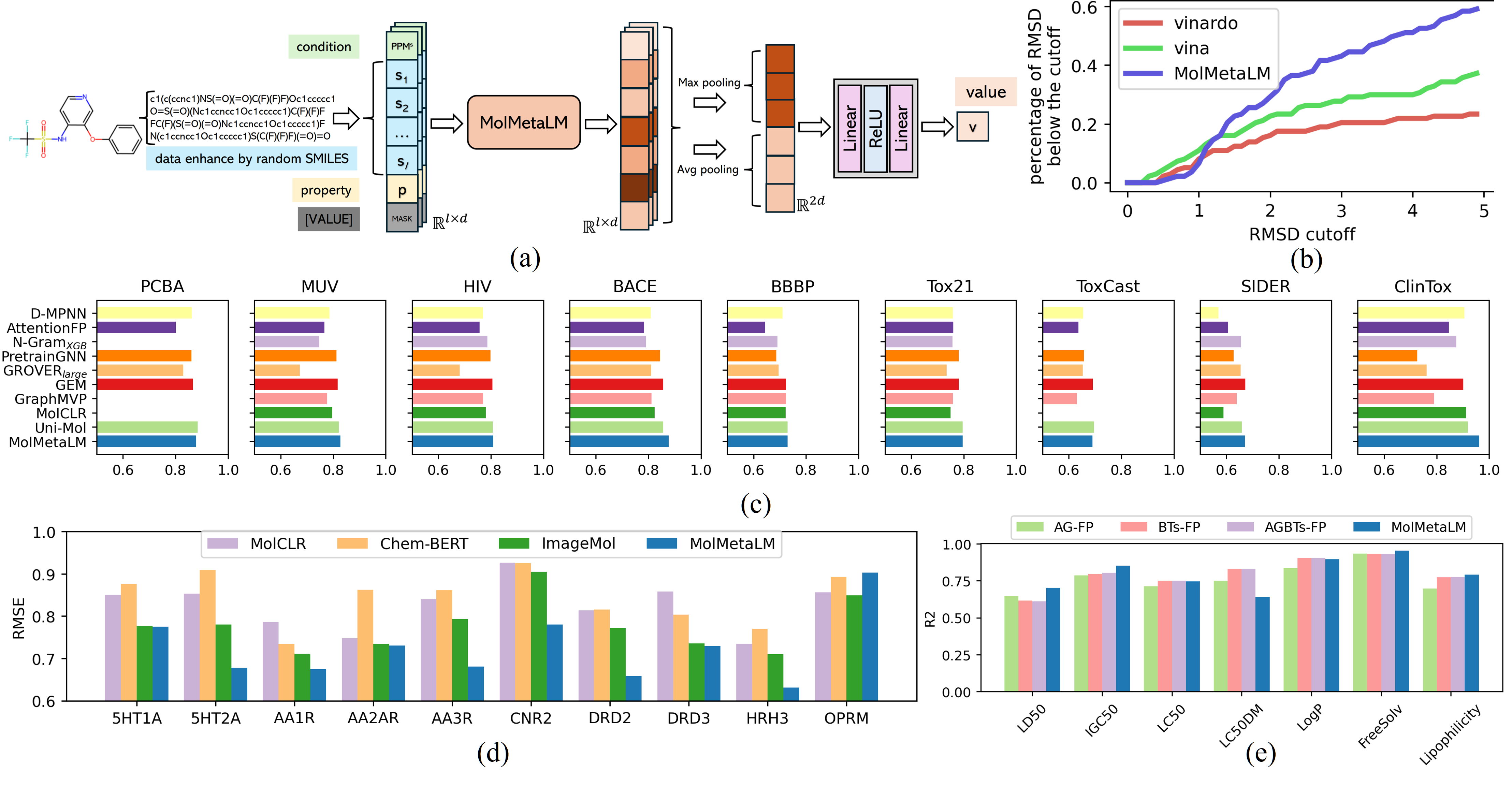}
    \caption{(a) The fine-tuned framework of MolMetaLM for molecular property prediction. First, SMILES are input into the pre-trained MolMetaLM to obtain the sequential molecular representation. Then a max-pooling and an avg-pooling are applied to extract the molecule-level features. Finally, the molecule-level features are fed into a FFN to generate the final predicted property value. (b) Blind docking results on the CASF-2016 test set. x-axis is the RMSD cutoff, y-axis is the percentage of ligand RMSD below the cutoff. These curves indicate how accurate these methods in docking ligands into the correct binding pocket and predicting the correct binding poses. (c) Classification results of MoleculeNet benchmark datasets. x-axis denotes different methods, y-axis is the macro AUROC results. (d) RMSEs of molecular activity prediction for the ten GPCR targets. (e) Regression results of comparing to AGBT's benchmark datasets. x-axis indicates 7 molecular property regression datasets, y-axis is R2 metrics defined as the squared Pearson correlations.}
    \label{all_res_prediction}
\end{figure*}

In addition to generation tasks, it is also impressive to fine-tune MolMetaLM to some specialized downstream prediction tasks (Figure. \ref{all_res_prediction} (a)). 

\noindent\textbf{Molecular property prediction.} We conduct experiments on six regression datasets \cite{chen2021algebraic} and nine classification datasets \cite{wu2018moleculenet} encompassing tasks such as molecular toxicity \cite{wu2018quantitative}, partition \cite{cheng2007computation}, and more. In the regression tasks, squared Pearson correlations (R2) are reported in Firugre. \ref{all_res_prediction} (e) and MolMetaLM outperforms AGBTs-FP \cite{chen2021algebraic}, which is a well-designed framework combining algebraic graph fingerprint and bidirectional transformer, in 4 out of 6 datasets. In the classification tasks, macro AUROCs are reported in Figure. \ref{all_res_prediction} (c) and MolMetaLM achieves the SOTA results in 7 out of 9 datasets. 

\noindent\textbf{Molecular activity prediction.} Molecular activity prediction is one of the fundamental tasks in drug discovery, which can effectively demonstrate the application potential of the model in the pharmaceutical industry. RMSEs of molecular activity prediction for the ten GPCR targets are reported in Figure. \ref{all_res_prediction} (d). MolMetaLM achieves the lowest RMSE in 9 out of the 10 datasets. 

\noindent\textbf{Molecular binding conformation prediction.} Molecular binding conformation prediction is an extremely challenging task because it requires the model to predict both the conformation of the molecule and the relative pose with respect to the whole protein. Compared to the AutoDock Vina \cite{trott2010autodock,eberhardt2021autodock} and Vinardo \cite{quiroga2016vinardo}, MolMetaLM demonstrates superior blind docking performance on the CASF-2016 test set when the ligand RMSD cutoff is over 1.4Å (Figure \ref{all_res_prediction} (b)). 

\subsection{Mechanism analysis} 

\begin{figure*}
    \centering
    \includegraphics[scale=0.43, trim=0 0 0 0]{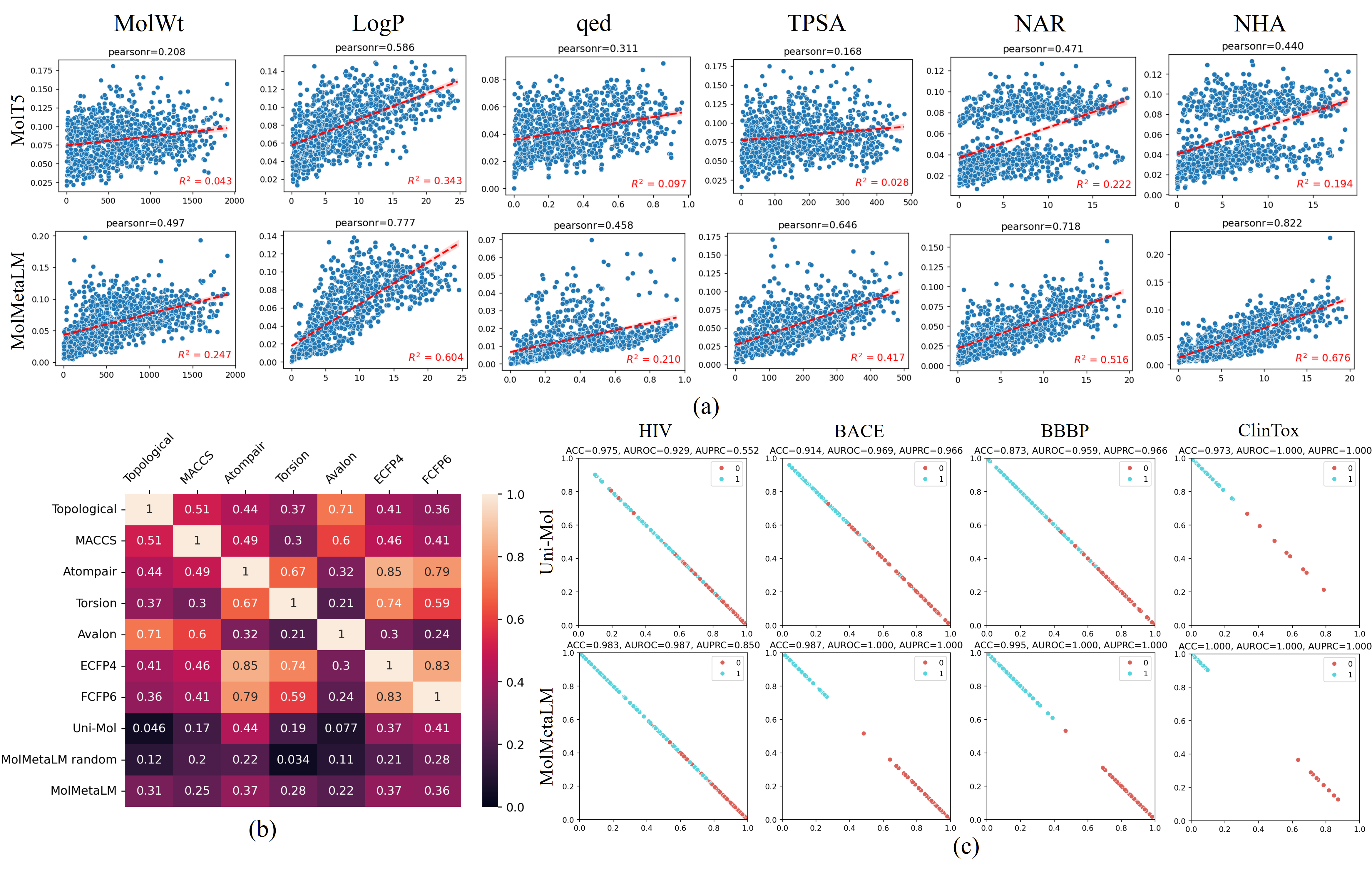}
    \caption{(a) Scatter plot of the numerical differences in the condition sequences versus the cosine distances of the constraint embedded vectors. x-axis denotes the absolute values of the numerical differences in the condition sequences, y-axis is the cosine distance between their embedded vectors. (b) Pearson correlation coefficient of similarities obtained by different molecular fingerprints or representations. The Pearson correlation coefficients (pearsonr) between them are shown in the title, and the coefficients of determination ($R^2$) of the fitted line are recorded in the bottom right corner of each figure. (c) Linear separable boundaries for molecular representations of Uni-Mol and MolMetaLM on four binary classification datasets. }
    \label{all_res_anysis}
\end{figure*}

\noindent\textbf{Sensitivity of numerical embedding analysis.} The key point of molecular generation is to enable the language model to accurately capture the conditional constraints in the input sequence, especially the numerical constraints, which requires the model to be sensitive to the numerical change of the condition sequence. MolMetaLM is trained on our designed molecular meta language with some special formats for the numerical condition values, which should help model better capture such numerical differences in conditions, compared with previous molecular language models. We randomly select 1,000 pairs of values from valid intervals for properties MolWt, LogP, qed, TPSA, NAR and NHA to generate the condition sequences, respectively. Then the condition sequences are fed into the language models to get the embedded vectors which are used for molecular generation by autoregressive prediction. After that, we compute the Pearson correlation coefficient (pearsonr) \cite{bravais1844analyse} between the absolute values of the differences of the 1,000 pairs of values and the cosine distance of their embedded vectors (Figure \ref{all_res_anysis} (a)). MolT5 \cite{edwards2022translation}, as a standard molecular language model, is used as the baseline method. We can find that for properties such as LogP that appear frequently in training corpus or literature, MolT5 can achieve a good numerical sensitivity of pearsonr almost 0.6, but for those rare or complex properties such as MolWt and TPSA, the numerical sensitivity is weak with pearsonr around or less than 0.2. In contrast, MolMetaLM shows significantly better numerical sensitivity, with pearsonr generally exceeding 0.6. This also proves that MolMetaLM can more accurately understand the numerical constraints in the conditions to achieve more accurate molecule generation, and explains the excellent generation performance of MolMetaLM in previous single-condition and multi-condition generation tasks (Figure \ref{all_res_generation}). 

\noindent\textbf{Robustness of molecular representation analysis.} Molecular fingerprints have already been proven to be very robust molecular representations and are widely used in drug discovery \cite{david2020molecular}. Analogously, representations obtained by molecular language models should also have some correlation with the traditional molecular fingerprints. We randomly select 1000 molecules outside the training set with different scaffolds, and use seven traditional molecular fingerprint methods represent these molecules and obtain the similarities between each pair of molecules. MolMetaLM's molecular representations are also obtained and we use cosine similarity to measure the similarities between molecules. Uni-Mol and a randomly initialized MolMetaLM (named MolMetaLM random) are used as baselines. The Pearson correlation coefficients between these similarities are shown in Figure \ref{all_res_anysis} (b). We can find that the fingerprints MACS, Atompair, ECFP4 and FCFP6 have a correlation of more than 0.2 even with the molecular representation obtained by a randomly initialized molecular language model. This is because they only encode some shallow features, such as statistical information of atoms or fragments, which can be captured easily at the token level by language models. On the contrary, for fingerprints Topological, Torsion and Avalon, the correlation is very low because they represent more complex molecular topological features or physicochemical information. Under the physicochemical knowledge-guided molecular meta language denoising pretraining, MolMetaLM improves the Pearson correlation with these three fingerprints to 0.31, 0.28 and 0.22 respectively, effectively empowering the complex information related to physicochemical properties into the molecular representation. For Uni-Mol, although it has better correlation on Atompair and FCFP6, it does not perform well on Topological, Torsion and Avalon fingerprints, which implies its limitations of embedding complex molecular patterns or features. In addition, we also use the four binary classification datasets HIV, BACE, BBBP and ClinTox to compare the linear separable boundaries of molecular representations from Uni-Mol and MolMetaLM without fine-tuning. Logistic regression models are employed to check the linear separable boundaries and the 2-dimension outputs are plotted in Figure \ref{all_res_anysis} (c). The accuracy (ACC), AUROC and AUPRC are reported at the top of each sub-figure. It is obvious that the molecular representations of MolMetaLM have nearly perfect linearly separable boundaries on almost all the four datasets, especially the AUROC and AUPRC of 1.0 on BACE, BBBP and ClinTox datasets. Uni-Mol has relatively fuzzy boundaries that requires some following fine-tuning to correct. 

\begin{figure*}
    \centering
    \includegraphics[scale=0.36, trim=0 0 0 0]{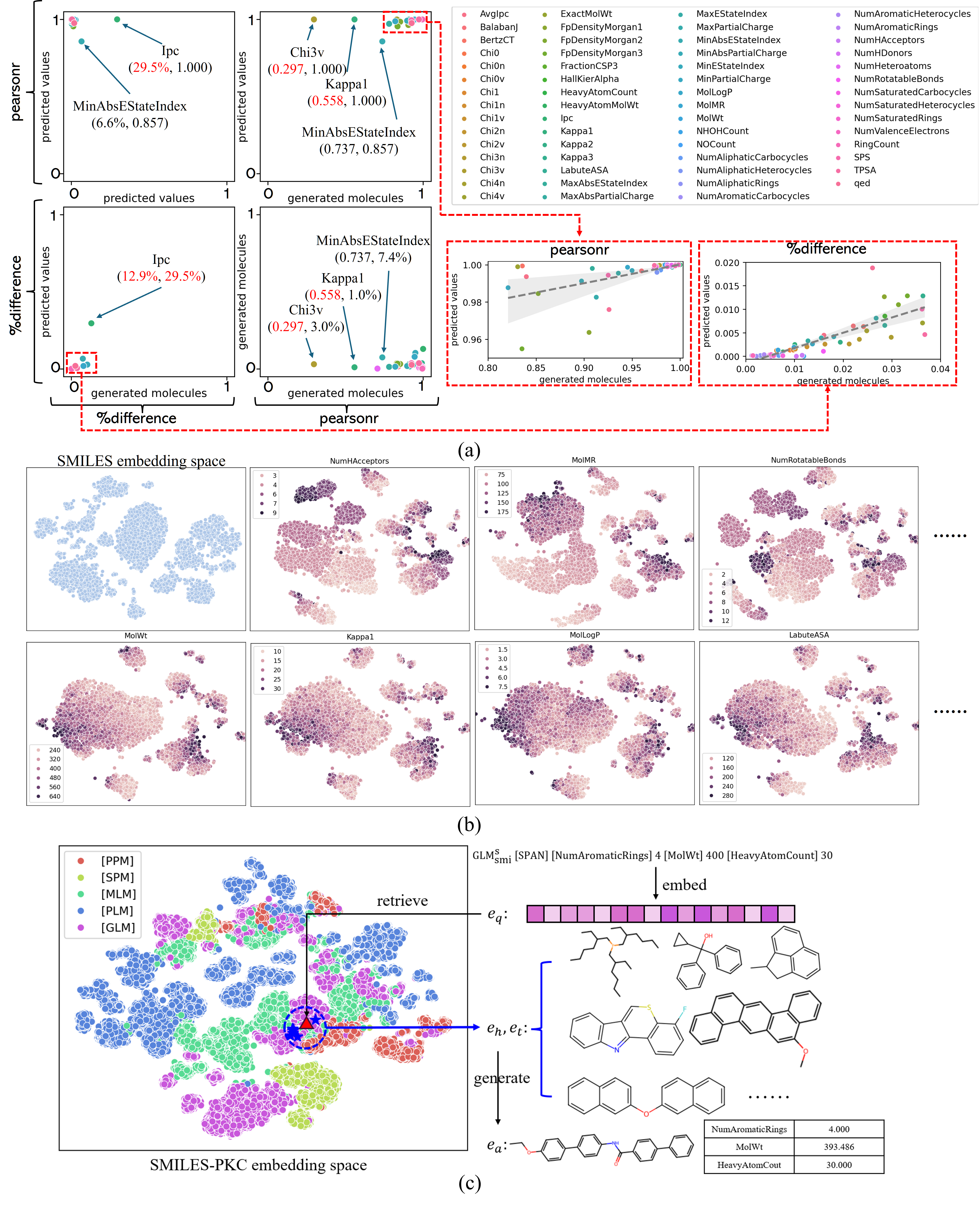}
    \caption{(a) Correlation between the performance of SMILES-based property prediction and property-based molecule generation. (b) Molecule embedding space and the variant of the space with the introduction of properties. (c) The analogical reasoning process of MolMetaLM. Generally speaking, during the pretraining process, MolMetaLM learns the embedding of all training samples and constructs the SMILES-PKC embedding space. During the inference process, MolMetaLM acquires the embedding $e_q$ of the query sequence and retrieves the relevant samples $(e_h,e_t)$ from the memorized training samples. Finally, the result is generated by integrating the retrieved samples and performing the analogical reasoning in the SMILES-PKC embedding space. }
    \label{all_res_embedding_space}
\end{figure*}

\noindent\textbf{Semantic understanding for properties and values analysis.} We conduct an experiment involving thousands of molecules that don't appear in the training set and let MolMetaLM perform both property prediction and property-based molecule generation. "[$\mathrm{PPM_m}$],S,[SEP],$p_1$,[VALUE],...,$p_k$,[VALUE]" and "[$\mathrm{GLM^s_{smi}}$],S,[SEP],$p_1$,$v_1$,...,$p_k$,$v_k$" are used as the input meta sequence. Then the generated sequences are transformed into property values (for molecular generation, property values are computed from the generated molecules using RDKit). We calculate the pearsonr and the percentage of value difference (\%difference) between the generated/predicted values $\{\hat{v}\}$ and ground truth values $\{v\}$ for each property. 

\begin{align}
\mathrm{pearsonr} &= \frac{\mathrm{cov}(\{\hat{v}\},\{v\})}{\sigma(\{\hat{v}\})\sigma(\{v\})} \\
\mathrm{\%difference} &= \mathrm{mean} \{\frac{|\hat{v}-v|}{|\hat{v}|+|v|}\}
\end{align}

\noindent where $\mathrm{cov}(\cdot)$ is the covariance function, $\sigma(\cdot)$ is the standard deviation function. As shown in Figure \ref{all_res_embedding_space}(a), MolMetaLM demonstrates exceptional semantic understanding for most of the properties in both molecular property prediction and property conditionally generation. However, there are some issues in property conditionally generation tasks of Chi3v, Kappa1, and MinAbsEStateIndex. The pearsonr values between the property values of the generated molecule and the conditions are only 0.297, 0.558, 0.737, respectively. For the property Ipc, both conditional molecule generation and molecular property prediction have significant errors, reaching 12.9\% and 29.5\% of \%difference, respectively. We further find that the lower pearsonr values for Chi3v, Kappa1 can be attributed to a common repetition problem encountered in generative models \cite{fu2021theoretical}. In some cases, MolMetaLM generates "C" recurrently in an infinite loop. Adding a repetition penalty to the decoding process does not provide a fundamental solution. This issue primarily arises when the model needs to generate molecules based on unseen property numerical constraints, indicating that MolMetaLM does not fully understand the continuous relationships between the values which are discretized into tokens. We hypothesize that more training data or non-discretized embedding of property value constraints would greatly alleviate this issue. For property Ipc (defined as the information for polynomial coefficients based on information theory \cite{nolte2017quantitative}), it has a very huge standard deviation ($>7\times 10^{28}$) and even is huge challenging for a regression model to fit well. In the future, language models need to pay more attention to semantically capturing the continuous information between numeric values, which may represent the most essential gap between advanced artificial intelligence and human intelligence.

\noindent\textbf{Embedding space of SMILES and physicochemical property analysis.} The essence of meta language model is to further refine and regularize the embedding of knowledge by language model. MolMetaLM will degenerate into a general MLM/GLM-based molecular LM by removing the property part of the sequence. The property part leads different embedding variants based on the original SMILES embedding space (Figure \ref{all_res_embedding_space} (b)). Each point denote a embedded molecule sample outside the training set dimensionally reduced by T-SNE \cite{van2008visualizing}. The original SMILES embedding space is obtained from "S" while the embedding variant is obtained from "[$\mathrm{PPM_s}$],S,[SEP],$p_1$,[VALUE]". We can see that with introduction of NHA and MolMR (Molecular Molar Refractivity), the embedding space tends to align with the corresponding property, forming a property-related embedding variant. In the embedding variant, molecules sharing similar properties are distributed more smoothly, which is conducive to the model's arrangement of the molecular latent space. Moreover, the space variants of related properties exhibit some associations. For example, molecules with higher molecular weights tend to have more complex topological structures (higher Kappa1 \cite{hall1991molecular}), more hydrophobic groups such as aromatic rings or long-chian alkyl (higher MolLogP), and larger accessible surface areas (higher LabuteASA \cite{labute2000widely}). For these related properties, similar correlations can also be found in their property-based embedding space variant (Figure \ref{all_res_embedding_space} (b)). This kind of physicochemical property-based embedding variant is regulated and described by our designed meta language, driving the model to better enhance the semantic relationships between physicochemical knowledge and the molecule itself. 

\noindent\textbf{Analogical reasoning on SMILES-PKC embedding space.} Meta language model can be regarded as introducing the description of knowledge in knowledge graph \cite{zhang2022multimodal} into language models to achieve a more refined modeling of knowledge. In turn, it makes it possible to interpret the mechanism of MolMetaLM by analogical reasoning. As shown in Figure \ref{all_res_embedding_space} (c), there are three steps: 1) embedding the input $<[\mathrm{SPAN}], p_1, v_1, ..., p_k, v_k>$ to $e_q$; 2) retrieving the relevant training samples and obtaining $e_h$ and $e_t$; 3) generate the missing entity (named as $e_a$) in the pair $(e_q, ?)$ based on the relevant entity pair $(e_h,e_t)$. MolMetaLM learns and memories many training molecular meta sequences to construct the Physicochemical Knowledge Conditioned SMILES embedding space (namely SMILES-PKC embedding space), where each point represents a sequence "$s_1,s_2,...,s_l, p_1,v_1,p_2,v_2,...,p_k,v_k$". Given a query sequence, MolMetaLM retrieves many relevant memorized sequences and generate the final results. Here, the SMILES-PKC embedding space is obtained by 100,000 training meta sequences from the training corpus and all embeddings are reduced to 2 dimensions using T-SNE \cite{van2008visualizing}. In general, we believe that the generative mechanism of all generative models can be interpreted in this framework, and potentially reflecting the thinking process of our human brain.

\section{Discussion}

We have designed a novel molecular meta language to describe and generate training sequences for pre-training molecular language models. The designed meta language paradigm consists of 18 basic tasks based on token-level denoising, sequence-level denoising, and order-level denoising. By introducing over 400 molecular physicochemical properties and fingerprint/conformation tasks, this paradigm can generate more than 10,000 pre-training tasks, effectively utilizing the training data. Compared with the original language model, the design pattern of meta language can drive the model to better understand the relationships between different parts of the input sequence, enhancing the logical reasoning and calculation ability of the language model.

We highlighted several improvements of MolMetaLM compared with existing molecular language models or pretraining frameworks. First, MolMetaLM achieves excellent performance across different generation tasks, including single/multiple property-based molecular generation, unconditional molecular generation, and molecular property optimization. Additionally, MolMetaLM achieves remarkable results in molecular property/affinity/conformation prediction tasks comparing to previous SOTA methods. Second, our designed molecular meta language is formatted as $k$ $<\mathrm{S},\mathrm{P},\mathrm{O}>$ triples, sharing the same $\mathrm{S}$ (SMILES). This meta language paradigm offers ease of extension and flexibility, allowing for the straightforward introducing of other molecule-related knowledge and tasks. Furthermore, MolMetaLM only needs SMILES as input, which is easily accessible and significantly broadens its potential applications. We will continue to leverage more data to explore the upper bound of semantic understanding at the language level. 

However, MolMetaLM also has some limitations. The first and most notable limitation is the natural lack of understanding of numerical values in language models. Since the language model needs to discretize numerical values into tokens for embedding, the natural relationships between the numerical values is lost. Language models can only explore the correlations between these values through a large amount of data training, which leads to potential errors in the model's understanding of certain unseen numerical values. One example is that values such as 1 and 10, which differ by a factor of 10 but differ by only one character in the discretized token space. We believe that there will be an elegant solution to this problem, which may also reveal some fundamental differences between current large language models and our human brain. Secondly, for some special inputs, MolMetaLM may encounter an issue of repeatedly generating the same segment/character and falling into an endless loop. We speculate that this issue is somewhat related to the previous one, and training with more data or using a larger model will alleviate it. ChatGPT, for example, with billions of parameters, has a lower probability of encountering this issue. In addition, this phenomenon may also be related to the Hallucination problem \cite{xu2024hallucination} in generative models, or that the repetitive generation problem is simply translated into the Hallucination phenomenon in the larger models with larger training data. The essence of the problem may still be that inadequate training data, causing the model to enter a state of "nonsense" when it encounters an input it doesn't understand. Finally, we acknowledge the training data of our model is limited and biased, that is, the molecules used for training are not uniformly distributed in the property space, which is biased to the human research preference. For an ideal physicochemical properties-guided molecular meta language model, we aspire for the training data to exhibit uniformly throughout the property space, enabling the model to equally understand the meaning of each property from all perspectives. One possible solution is to allow the model to generate additional molecules as new training samples to balance the sampling of molecules in different property spaces. We will focus on this in the future.

\section{Methods}

\subsection{Generative language model}

A Generative Language Model (GLM) is a kind of language model designed to generate target sequences based on given source sequences. It works by iteratively predicting the probability of the next token based on the previous tokens. Let $X^s=[x^s_1,x^s_2,...,x^s_{l_s}], X^t=[x^t_1,x^t_2,...,x^t_{l_t}]$ are the source and target sequences of training samples. For a GLM $\mathcal{G}_\theta$, the commonly used loss function is as follows:

\begin{equation}
\mathcal{L} = \arg \min_\theta \sum_{i=1}^{l_t} -x^t_i \mathrm{log} (\mathcal{G}_\theta (X^s,[x^t_1,x^t_2,...,x^t_{i-1}]))
\end{equation}

\noindent where $l_s$ and $l_t$ represent the length of the source and target sequences, $x^s_i,x^t_i \in \mathbb{R}^{1 \times m}$ are one-hot vectors indicating the real tokens in the source or target sequences, $m$ denotes the number of tokens in the dictionary, $\mathcal{G}_\theta (X^s,[x^t_1,x^t_2,...,x^t_{i-1}]) \in \mathbb{R}^{m \times 1}$ represents the predicted probability distribution of token ${i}$, given the known source sequence $X^s$ and the previous tokens $[x^t_1,x^t_2,...,x^t_{i-1}]$. GLM provides us an excellent learning framework that can easily integrate various pre-training tasks or strategies. For the given sequence $[x_1,x_2,...,x_l]$: randomly masking some tokens and setting $X^s=[\mathrm{MASK},x_2,x_3,\mathrm{MASK},x_5,...,x_n], X^t=[x_1,x_4,...]$, the MLM task can be implemented. Shuffling the sequence and let the shuffled sequence as $X^s$, the Permutation Language Model (PLM) task can be implemented; Generating the original sequence, or setting $X^s=[x_1,x_2,...,x_{l_s}], X^t=[x_{l_s+1},x_{l_s+2},...,x_{l}]$, we can derive the GLM task. The current large language model such as GPT \cite{brown2020language}, Llama \cite{touvron2023llama}, ChtGLM \cite{zeng2022glm} all use this GLM framework, and it is one of the most promising neural network architectures for artificial general intelligence \cite{ge2024openagi}. In our study, we choose Llama as the backbone of MolMetaLM, which incorporates the RMSNorm \cite{zhang2019root}, SwiGLU \cite{shazeer2020glu}, and rotary position embedding \cite{su2024roformer} into the Transformer block, as shown in Equations. \ref{formula_llama_s} to \ref{formula_llama_e}.

\begin{align}
\label{formula_llama_s} X &= \mathrm{RMSN}(X) \\
Q^i &= X W^i_Q, K^i = X W^i_K, V^i = X W^i_V \\
H^i &= \frac{\mathrm{softmax}((\mathcal{R}_m Q^i)(\mathcal{R}_m K^i)^\mathrm{T})V^i}{\sqrt{d_k}} \\
X &= X+\mathrm{concat}(H^1,H^2,...)W_Z \\
\label{formula_llama_e} X &= X + \mathrm{SwiGLU}(\mathrm{RMSN}(X))
\end{align} 

\noindent where $X \in \mathbb{R}^{l \times d}$ is the vector representation of the input sequence, $W^i_Q,W^i_K,W^i_V \in \mathbb{R}^{d \times d_k},W_Z \in \mathbb{R}^{d_k \times d}$ are learnable weight matrices, $\mathcal{R}_m \in \mathbb{R}^{d \times d}$ represents the rotary matrix, as defined in \cite{su2024roformer}, $\mathrm{RMSN}(\cdot)$, i.e. Root Mean Square Layer Normalization, denotes the layer normalization with zero expectation and bias, $\mathrm{SwiGLU}(\cdot)$ represents the GLU-based FFN (Feed Forward Network) with activation function $\mathrm{Swish}(\cdot)$ \cite{ramachandran2017searching}.

\subsection{Denoising pre-training strategies}

For a GLM, the most important thing is the definition of training samples, i.e., the design of denoising pre-training strategies. For the GPT or other general GLMs, we can directly extract unstructured natural language text for training. However, for specialized GLMs or meta LMs, the utilization of manually designed structured training samples can be more effective in driving the model to understand domain-specific knowledge and achieve a more comprehensive representation. In the field of molecules, the most basic and nature expertise is the physical and chemical properties, such as molecule weight, fraction of SP3, the count of hydrogen acceptors/donors. To incorporate these physicochemical knowledge into the training samples, we develop a novel pre-training paradigm for molecules. In this paradigm, first, the training samples of molecules are viewed as a mixture of $k$ $<\mathrm{S},\mathrm{P},\mathrm{O}>$ triples that share the same $\mathrm{S}$. We call it as molecular meta language because it is essentially a well-defined molecular meta language paradigm. Next, by introducing the token-level, sequence-level, order-level noises to each element of the triples, and combining hundreds of molecular physicochemical properties, this paradigm can spawn tens of thousands of denoising MLM/GLM/PLM pre-training tasks.

\subsection{Fingerprint and conformation prediction tasks for pre-training}

Both fingerprints and conformations are descriptors of the global properties of molecules, which integrates many molecular physical and chemical properties. By introducing the fingerprint and conformation prediction tasks into the pre-training process, we can greatly improve the model's ability to capture global molecular features. For the fingerprint prediction task, we set the property names and property values as the type and bit values of the fingerprint. Specifically, we employ five fingerprints, namely MACCS \cite{durant2002reoptimization}, Topological \cite{nilakantan1987topological}, FCFP, ECFP \cite{rogers2010extended}, and Avalon \cite{gedeck2006qsar} to extract the global features of molecules. To ensure efficient training, the length of all fingerprints is set to 176. For the conformation prediction task, the property names are derived from the atom names in the input SMILES, following the same sequential order. The property values correspond to the coordinates of the corresponding atoms. The most challenging thing is how to represent the conformation as a sequence with an acceptable length while satisfying the translation and rotation invariance. We know that only the local coordinates are not changed while we translate or rotate the whole system. Therefore, we develop a novel translation-rotation-invariance sequential representation for conformation. In particular, each position of the atom is represented as local coordinates based on a local frame constructed by its preceding three atoms. For the construction of the local frame, based on the distance between 2 atoms, inner angle between 3 atoms, and dihedral angle between 4 atoms, we give 3 types of local frame implementations. Take one of them as an example, for 4 atoms $P_1,P_2,P_3,P_4$ that are continuous on SMILES representation, the coordinates of $P_4$ can be determined by the distance $d$ of $P_4-P_3$, the inner angle $\alpha$ of $P_2-P_3-P_4$, and the dihedral angle $\beta$ between the planes of $P_1-P_2-P_3$ and $P_4-P_2-P_3$. Therefore, the local coordinates of $P_4$ can be denoted as ($\alpha,\beta,d$).

\subsection{Implementation details}

During the training process, we utilize all the molecules available in PubChem \cite{wang2009pubchem} (99.95\% for training, 0.05\% for validating), which comprises over 110 million molecules. We use RDKit \cite{landrum2013rdkit} to extract the physicochemical properties and all properties defined in "rdkit.Chem.Descriptors" are used in our pre-training, containing 402 properties in total. MolMetaLM has 12 transformer blocks with 12 attention heads and a hidden size of 768. The FFN expansion size is set to 2,560. All weights are open sourced on Huggingface \cite{wolf2019huggingface} at \url{https://huggingface.co/wudejian789/MolMetaLM-base}. During the training process, MolMetaLM is trained using 4 $\times$ NVIDIA Tesla A100 (40GB) GPUs. The training is conducted for a total of 1,000,000 steps with batch size of 256 and learning rate of 0.0001. 

\section{Data availability}

The datasets used in MolMetaLM can be found at the following links: over 110 million molecular SMILES used for pre-training MolMetaLM, \url{https://ftp.ncbi.nlm.nih.gov/pubchem/Compound/Extras/CID-SMILES.gz}; 6 molecular property regression datasets from AGBT, \url{https://weilab.math.msu.edu/Database/}; 9 molecular property classification datasets from MoleculeNet, \url{https://bioos-hermite-beijing.tos-cn-beijing.volces.com/unimol_data/finetune/molecular_property_prediction.tar.gz}, or the raw MoleculeNet dataset, \url{https://github.com/deepchem/deepchem}; 10 GPCR-related molecular activity datasets, \url{https://drive.google.com/file/d/1HVHrxJfW16-5uxQ-7DxgQTxroXxeFDcQ/view}; Molecular binding conformation datasets, \url{http://www.pdbbind.org.cn/download.php} for PDBbind General set v.2020, \url{http://www.pdbbind.org.cn/casf.php} for CASF-2016. 

\section{Code availability}

All source code involving designed model, pre-training framework and experiments are available at \url{https://github.com/CSUBioGroup/MolMetaLM}. 



\backmatter

\bmhead{Acknowledgements}

This work is supported in part by the National Natural Science Foundation of China under Grant (No.62225209 to M.L.), Hunan Postgraduate Research and Innovation Project (No.1053320213431 to Y.W.).

\section*{Declarations}

The authors declare no competing interests.


\bibliography{sn-bibliography}

\end{document}